# Coherence interpretation of the Hong-Ou-Mandel effect

Byoung S. Ham

School of Electrical Engineering and Computer Science, Gwangju Institute of Science and Technology
123 Chumdangwagi-ro, Buk-gu, Gwangju 61005, S. Korea
(Submitted on November 21, 2022; bham@gist.ac.kr)

**Abstract**
Two-photon intensity correlation of the Hong-Ou-Mandel (HOM) effect has been intensively studied over the last several decades for one of the most interesting quantum features. According to the particle nature of quantum mechanics, indistinguishable photon characteristics interacting on a beam splitter are the prerequisite of the photon bunching phenomenon. Here, a coherence approach based on the wave nature of a photon is used to interpret HOM effect based on entangled photon pairs. As a result, a complete solution of the HOM effect is derived from the coherence approach for the indistinguishable photon characteristics in a deterministic way without violation of quantum mechanics. Thus, HOM effect is now perfectly understood as a relative phase relation between paired photons in an interferometric system, where the HOM dip with no interference fringe is due to ensemble decoherence of all interacting photon pairs.

**Introduction**

In 1987, a seminal paper [1] was published by Hong, Ou, and Mandel to demonstrate a photon bunching phenomenon on a beam splitter (BS) using paired entangled photons generated from a spontaneous parametric down conversion (SPDC) process [2,3]. Since then, the photon bunching effect on a BS has been intensively studied in the name of the Hong-Ou-Mandel (HOM) effect as one of the most interesting quantum phenomena [4-14]. The physics of the HOM effect has been explained as destructive quantum interference between finding probabilities of two interacting photons on a BS according to the particle nature of quantum mechanics [15]. Even though an absolute phase is not allowed to a single photon by the Copenhagen interpretation, allowing a definite relative phase between paired photons does not violate quantum mechanics. Thus, an unspoken relative phase has been used for the analysis of the HOM effect to come up with the BS matrix representation based on coherence optics [1,15]. Although this in-phase relation between paired photons is sustainable in general quantum mechanics, such an absolute phase relation has never been discussed in the main stream of quantum information. Some observed HOM effects for completely independent molecules [6] and photons [10] are thus more curious on this quantum mechanism. Thus far, all varieties of HOM experiments have been conducted to testify indistinguishable particle natures with no phase information [4-14]. Here, the HOM effect is investigate to determine the definite phase relation between paired photons using the wave nature of a photon. For this, a pure coherence approach is applied to the BS system interacting with entangled photon pairs generated from the SPDC process.

Recently, a coherence approach has been conducted to understand the HOM effect using coherent photons, where a relative phase between paired photons is achieved in a Mach-Zehnder interferometer [17,18]. In this coherence interpretation, a solution of the relative phase relation has been derived analytically, where the indistinguishable photon characteristics on a BS are coherently understood for the sum of many wave interferences. In this coherence approach, the two-photon correlation can be appeared classically or nonclassically, depending on the relative phase [17]. The observed HOM dip with no interference fringes has also been investigated [18]. Here, a complete solution of the HOM effect is derived on a BS interacting with entangled photon pairs generated from SPDC process of $\chi^{(2)}$ nonlinear optics [2,3]. From this analytical solution, the HOM effect is demonstrated for the phase-dependent quantum feature in a deterministic way. Thus, the coherence approach can also give a solution to the two-mode HOM effect based on independent light sources [6,10] via phase manipulation of the incoherent photons in an optical cavity [19]. The coherence interpretation of indistinguishable photons has been sought differently in an interferometric system [20]. The HOM dip with no interference fringe is also numerically confirmed as a direct result of ensemble dephasing. As a result, the present work may give an answer to the unanswered question of the quantum mystery in quantum computing [21] and quantum interface [22] based on Bell measurements via the HOM effect.

**Results**

Figure 1 shows schematic of typical HOM experiments [1], whose paired photons ('s' and 'i') are generated from the SPDC process [2,3]. The HOM effect has also been tested using thermal and coherent photons from independent and individual light sources [6,10]. In ref. [6], a coherent pump light may excite phase coherence between simultaneously and spontaneously generated photons from individual molecules. In ref. [10], such a coherence relation between independent photons can also be excited through a pair of identical optical cavities. Unlike coherent photons governed by Poisson statistics, each SPDC-generated photon pair has an opposite detuning relation as shown in Fig. 1(a), whose frequency detuning ($\pm \delta f_j$) from the center frequency $f_0$ is symmetric to satisfy the energy conservation law. This symmetric and opposite frequency relation between the signal and idler photons in SPDC process can be precisely provided by a narrow-linewidth pump laser ($f_p$) [2,3,12]. Entanglement is normally achieved via spatial mixing of the signal ($f_s$) and idler ($f_i$) photons, as shown in Fig. 1(b) [2,3]. Although each photon pair of SPDC has a random frequency detuning ($\pm \delta f_j$) within the spectral distribution in Fig. 1(a), the sum frequency of the paired photons must be fixed at the pump frequency according to the phase matching condition of $\chi^{(2)}$ nonlinear optics via energy and momentum conservation laws: $f_p = f_s + f_i$, $\mathbf{k}_p = \mathbf{k}_s + \mathbf{k}_i$ [2,3]. For degenerate SPDC photon pairs in Fig. 1(a), thus, $f_0 = \frac{1}{2} f_p$ is satisfied. The frequency stability of the pump photon $f_p$ ($= 2f_0$) plays an essential role to the visibility of photon bunching [1,12]. Due to the phase matching in $\chi^{(2)}$ nonlinear optics, a relative phase relation between the paired photons must be preserved in the coherence interpretation [3,23].

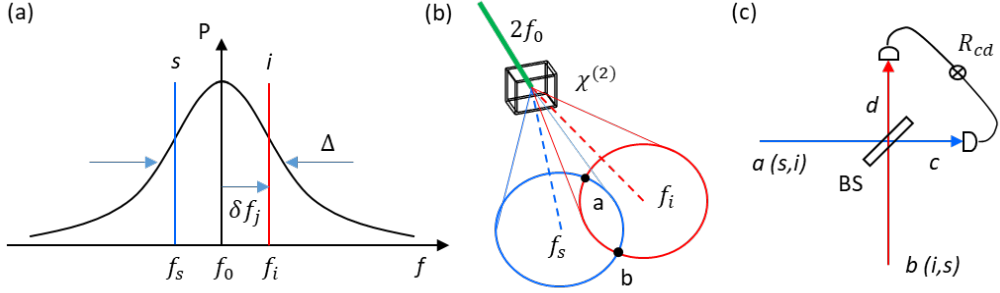

**Fig. 1.** Schematic of an entangled photon-pair based HOM effect. (a) SPDC-generated pPhoton pair. (b) SPDC-based entanglement. (c) HOM measurements. BS: nonpolarizing balanced beam splitter. s (i): signal (idler) photon.

In the present coherence approach, the SPDC-generated signal and idler photons are now represented by electric fields $E_s$ and $E_i$, respectively. The corresponding amplitude and phase of a single photon are $E_0$ and $\Delta_j$ ($= \delta f_j \tau$), where the subscript $j$ stands for an arbitrary $j^{th}$ photon pair in the SPDC process. Here, $\tau$ is the arrival-time delay of the idler photon on the BS with respect to the signal. For the symmetric detuning between the paired photons at $f_0 \pm \delta f_j$, the detuning $\delta f_j$ of the $j^{th}$ photon pair must be random (see Fig. 1(a)). Most importantly, the signal and idler photons picked up at superposed positions 'a' and 'b' in Fig. 1(b) are not determined, satisfying the entanglement relation in the particle nature of quantum mechanics: $|\Psi\rangle = \frac{1}{\sqrt{2}} \left( |signal\rangle_a |idler\rangle_b + e^{i\theta} |idler\rangle_a |signal\rangle_b \right)$; $\theta$ is the relative phase between two correlated states. Thus, the photon from the path 'a' in Fig. 1(a) can be either a signal ($f_s$) or idler ($f_i$). This is exactly the same as the $\pm \delta f_j$ relation. Likewise, the same randomness is satisfied for path 'b'. This is the original concept of the particle nature-based photon indistinguishability, resulting in uniform intensities in both output ports. Thus, any photon pair between paths 'a' and 'b' in Fig. 1(c) satisfies the entanglement relation $|\Psi\rangle$ on the BS [23]. We now derive the relative phase relation between the paired entangled photons for the HOM effect.

For the paired photons incident on the BS in Fig. 1(c), the system's basis must be related with the relative phase of the paired photons to come up with the BS matrix representation derived from pure coherence optics [15-17]. With a definite phase relation, Fig. 1(c) is equivalent to a typical double-slit system having two-input and two-output modes [24]. For the paired photons satisfying $|\Psi\rangle$, thus, we simply extend the same relative phase in each pair to all pairs. As a result, the following coherence equation can be straightforwardly obtained using the matrix representation of a BS [16] for any $j^{th}$ photon pair of $|\Psi\rangle$ mentioned above (see the Supplementary Materials):

$$\begin{bmatrix} E_c \\ E_d \end{bmatrix}_j = \frac{1}{\sqrt{2}} \begin{bmatrix} 1 & i \\ i & 1 \end{bmatrix} \begin{bmatrix} E_a \\ E_b \end{bmatrix}_j$$
$$= \frac{E_0}{\sqrt{2}} e^{i(kx-2\pi f_0 t)} e^{-i\Delta_j} \begin{bmatrix} 1 & i \\ i & 1 \end{bmatrix} \begin{bmatrix} 1 \\ e^{i(2\Delta_j+\theta_0)} \end{bmatrix}, \quad (1)$$

where $\theta_0$ is the assumed inherent phase shift of the idler photon with respect to the signal. In Eq. (1), the signal and idler photons correlate with input paths 'a' and 'b', respectively, indicating the first path-photon correlation term in $|\Psi\rangle$. For the second path-photon correlation term, the path choices are simply swapped for the same photon, resulting in the element swapping in the last matrix due to frequency swapping ($\pm\delta f_j$) in Fig. 1(a). Thus, the symmetric detuning relation in Fig. 1(a) for the full bandwidth implies the entanglement relation in $|\Psi\rangle$. Considering the hyper-THz SPDC bandwidth with respect to a sub-GHz linewidth of a pump laser [2,3,12], the center frequency $f_0$ is considered as a fixed parameter for all photon pairs. Thus, the sharpness of $f_0$ determines the visibility of HOM effects [1,11,12]. From Eq. (1), the amplitudes of the $j^{th}$ output photons in Fig. 1(c) are rewritten as:

$$(E_c)_j = \frac{E_0}{\sqrt{2}} e^{i(kx-2\pi f_0 t-\Delta_j)} \left(1 + i e^{i(2\Delta_j+\theta_0)}\right), \quad (2)$$
$$(E_d)_j = \frac{iE_0}{\sqrt{2}} e^{i(kx-2\pi f_0 t-\Delta_j)} \left(1 - i e^{i(2\Delta_j+\theta_0)}\right). \quad (3)$$

The corresponding output intensities are as follows:

$$(I_c)_j = I_0 \left(1 - \sin(2\Delta_j + \theta_0)\right), \quad (4)$$
$$(I_d)_j = I_0 \left(1 + \sin(2\Delta_j + \theta_0)\right). \quad (5)$$

For the second path-photon correlation in $|\Psi\rangle$, Eqs. (4) and (5) are simply swapped. For the detuning swapping in $\pm\delta f_j$, Eqs. (4) and (5) are also swapped. Satisfying entanglement relation in $|\Psi\rangle$, thus, the sum of intensities for full bandwidth distributed photon pairs results in self-cancellation of the sine terms. In other words, the spectral swapping between the signal and idler photons corresponds to the sign change in the sine terms of Eqs. (4) and (5) (see the Supplementary Materials): $(I_c)'_j = I_0\left(1 + \sin(2\Delta_j + \theta)\right)$; $(I_d)'_j = I_0\left(1 - \sin(2\Delta_j + \theta)\right)$. These primed intensities represent the second path-photon correlation term in $|\Psi\rangle$. Thus, time-averaged measurements in both output ports in Fig. 1(c) result in a uniform intensity:

$$\langle \overline{I_c} \rangle = \frac{1}{2N} \sum_{j=1}^{N} \left[(I_c)_j + (I_c)'_j\right] = I_0, \quad (6)$$
$$\langle \overline{I_d} \rangle = \frac{1}{N} \sum_{j=1}^{N} \left[(I_d)_j + (I_d)'_j\right] = I_0. \quad (7)$$

The linear superposition between two path-photon correlation terms in $|\Psi\rangle$ also satisfies Eqs. (6) and (7), where the resultant uniform intensity $I_0$ proves the first validity of the present coherence approach for the HOM effect. The non-coherence feature of indistinguishability in conventional understanding of particle nature-based quantum mechanics is now replaced by the coherence feature of the basis randomness. In Fig. 1(c), the system basis of the BS with two mode photons is either 0 or $\pi$, resulting in Eqs. (4) and (5). Swapping of the bases results in the primed intensities in Eqs. (6) and (7). In that sense, the HOM observation in refs. [6] and [10] intrigues a fundamental question of quantumness in the HOM effect. Entanglement equivalence between $|\Psi\rangle$ and the detuning swapping in Fig. 1(a) may give us a clue for the HOM effect. To answer these questions, we now analytically derive the value of $\theta_0$ in Eq. (1).

For the second-order intensity correlation $R_{cd}(\tau)$ between individually measured two output photons in Fig. 1(c), the HOM effect relates to the average of coincidence detections for all independently measured paired photons. The sign change in Eqs. (4) and (5) represents swapping between the signal and idler photons in Fig. 1(a), resulting in detection randomness in both detectors. However, this randomness between paired photons with respect to the input ports does not affect the intensity product of individually measured output photons, $(I_c)_j (I_d)_j$ or $(I_c)'_j (I_d)'_j$, due to the path-photon correlation in $|\Psi\rangle$, resulting in $R_{cd}(\tau) = I_c(t)I_d(t+\tau) = I'_c(t)I'_d(t+\tau)$. For $\langle R_{cd}(\tau)\rangle$, $\Delta_j$ in Eqs. (4) and (5) is severely affected by $\tau$ due to $\delta f_j \tau$. The common harmonic oscillation terms are already taken out in Eqs. (2) and (3). Thus, the τ-dependent second-order intensity correlation is represented by:

$$\langle R_{cd}(\tau) \rangle = \frac{1}{N} \langle \sum_{j=1}^{N}(I_c)_j \rangle \langle \sum_{j=1}^{N}(I_d(\tau))_j \rangle$$
$$= \langle I_0^2 \rangle \langle \sum_j^N cos^2(\delta f_j \tau + \theta_0) \rangle. \quad (8)$$

At $\tau = 0$, the HOM effect of $\langle R_{cd}(0) \rangle = 0$ is satisfied for all photon pairs in Fig. 1(a). The only mathematical condition for this is $\theta_0 = \pm \frac{\pi}{2}$, regardless of the SPDC bandwidth $\Delta$, because $\delta f_j \tau = 0$. This is the quintessence of the present coherence approach for the HOM effect.

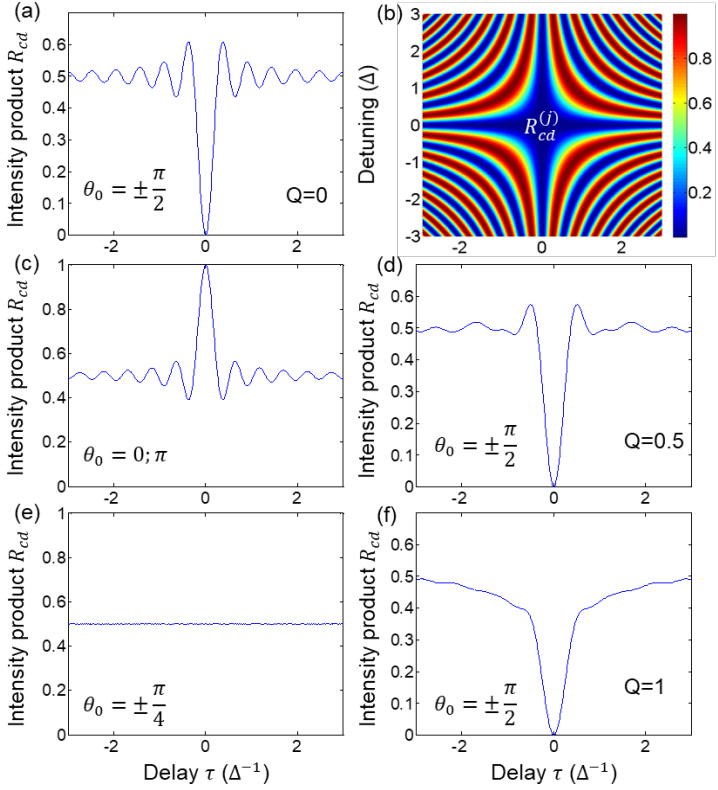

**Fig. 2.** Numerical calculations for Eqs. (6) and (7). (a) and (b) $\theta_0 = \pm \frac{\pi}{2}$ and $Q = 0$. (c) $\theta_0 = 0; \pi$ and $Q = 0$. (d) $\theta_0 = \pm \frac{\pi}{2}$ and $Q = 0.5$. (e) $\theta_0 = \pm \frac{\pi}{4}$ and $Q = 0$. (f) $\theta_0 = \pm \frac{\pi}{2}$ and $Q = 1$. Q is phase fluctuation factor. The unit of $R_{cd}$ is $I_0^2/2$.

Figure 2 shows numerical calculations for Eqs. (4) and (5) as a functions of the delay $\tau$ and detuning bandwidth $\Delta$ in Fig. 1. The top panels of Fig. 2 are for the solution of the HOM effect, where the phase difference between the paired input photons is set for $\theta_0 = \pm \pi/2$. Due to some experimental conditions, the photon pair may have some random phase fluctuation factor Q. This factor Q can be caused by SPDC-generated photon-pair collection efficiency due to imperfect spatial mode conditions as well as the pump laser linewidth. For Q-based calculations, a weighted random number $(1 - Q)$ is multiplied to $\Delta_j$ for comparison purposes. The left column of Fig. 2 shows the $\theta_0$-dependent intensity products for an ideal case with $Q = 0$, where Figs. 2(c) and (e) show a perfect thermal and classical photon cases, respectively. Figure 2(b) shows the original data of Fig. 2(a) for all detuned photon pairs representing the coherence feature of individual HOM effects. Figure 2(a) is the average of all detuned pairs in Fig. 2(b) with respect to τ. Figure 2(d) and (f) show the Q-factor dependent intensity products compared to Fig. 2(a), where the sideband wiggles become disappeared as Q factor increases. This type of dephasing has already been experimentally observed [1], where spectral filtering of SPDC photons recovers the coherence feature, resulting in sideband wiggles [7, 12]. Thus, the present coherence approach is well supported by numerical calculations for the analytical solutions.

**Discussion**

As $\tau$ increases, the cosine term in Eq. (8) increases, too, whose gradient depends on $\delta f_j \tau$. Due to many different photon pairs with $\delta f_j$, the ensemble average shows no $\lambda$-dependent fringe as shown in Fig. 2. Instead, slow degradation of the HOM effect is resulted as $\tau$ increases until reaches at $\langle R_{cd}(\tau) \rangle = \langle I_0^2 \rangle / 2$. Here, $\langle I_0^2 \rangle / 2$ is the lower bound of the classical physics in Fig. 2(e), representing incoherent and individual photons [17]. These are the second validity of the present coherence approach for the HOM effect. If the SPDC bandwidth gets narrowed, then wiggles appear in both sides of the HOM dip across $\tau = 0$, as shown in Figs. 2(a), (d), and (f) [12,18]. This wiggles are the indirect proof of the present coherence approach caused by the ensemble coherence. The SPDC spectral filtering in actual experiments results in reducing the Q factor. As shown in the left column of Fig. 2, the resultant relative phase $\pm \frac{\pi}{2}$ between paired photons must be an inherent property of SPDC-generated entangled photon pairs. Such a coherence relation with a definite relative phase can also be created even between independent light sources if identical optical cavities are used [10]. The present analysis of the HOM effect with the relative phase relation between all paired photons is unprecedented and cannot be derived from the conventional particle nature-based quantum mechanics [15].

**Conclusion**

One of the most important quantum features was analyzed for the HOM effect based on entangled photon pairs using the wave nature of quantum mechanics. Unlike conventional understanding of the phase-independent particle nature, the present coherence approach was perfectly analyzed for the observed uniform local intensities and photon bunching phenomenon of the HOM effect. Most of all, a definite phase relation between entangled photons was derived analytically as a critical condition for the photon bunching. Thus, the quantum mystery of the photon bunching was unveiled for the coherence relation satisfying indistinguishability for the second-order intensity correlations. The observed HOM dip with no interference fringes was also clearly explained by the ensemble decoherence. As a result, a complete understanding on two-photon correlation of the HOM effect was deterministically established.

**Declarations**
**Methods:** Not applicable.
Ethical Approval and Consent to participate: Not applicable.

**Availability of data and materials**
Not applicable. For requests relating to the paper, please contact the first author.
**Competing interests**
The author declares no competing interests.
**Funding:** This work was supported by the ICT R&D program of MSIT/IITP (2022-2021-0-01810), development of elemental technologies for ultrasecure quantum internet.
**Author's contribution:** BSH conceived the idea and wrote the paper.